# Single-shot, full characterization of the spatial wavefunction of light fields via Stokes tomography

**Bing-Shi Yu[1], Hai-Jun Wu[1], Chun-Yu Li[1], Jia-Qi Jiang[1], Bo Zhao[1], Carmelo Rosales-Guzmán[2], Bao-Sen Shi[3], and Zhi-Han Zhu[1*]**

[1] Wang Da-Heng Center, Heilongjiang Key Laboratory of Quantum Control, Harbin University of Science and Technology, Harbin 150080, China
[2] Centro de Investigaciones en Óptica, A.C., Loma del Bosque 115, Colonia Lomas del Campestre, 37150 León, Gto., Mexico
[3] CAS Key Laboratory of Quantum Information, University of Science and Technology of China, Hefei, 230026, China
* e-mail: zhuzhihan@hrbust.edu.cn

**Abstract**: Since the diffraction behavior of a light field is fully determined by its spatial wavefunction, i.e., its spatial complex amplitude (SCA) distribution, full SCA characterization plays a vital role in modern optics from both the fundamental and applied aspects. In this work, we present a novel 'complex-amplitude profiler' based on spatial Stokes tomography with the capability to fully determine the SCA of a light field in a single shot with high precision and resolution. The SCA slice observed at any propagation plane provides complete information about the light field, thus allowing us to further retrieve the complete beam structure in 3 dimensions space, as well as the exact modal constitution in terms of spatial degrees of freedom. The principle demonstrated here provides an important advancement for the full characterization of light beams with a broad spectrum of potential applications in various areas of optics, especially for the growing field of structured light.

## Introduction

The term 'wavefunction' usually refers to a complex-valued function in modern quantum theory that uses to completely describe a field entity and predict its dynamical behavior [1,2]. For a light field propagating in free space or in a preknown optical system, once its spatial wavefunction is determined at a certain propagation plane, the whole diffractive propagation can be known exactly. This function is commonly known as the spatial complex amplitude (SCA) [3]. Therefore, SCA characterization plays a vital role in modern optics for both fundamental studies and applications. The experimental determination of an SCA requires two complementary measurements to record its modulus and argument (or phase) distributions. The modulus distribution describes the amplitude structure of the light field and is an observable (or rather, its absolute square) that can be directly recorded via a variety of modern photoelectric detectors, such as a charge-coupled device (CCD). The argument distribution, which carries the wavefront structure information, cannot be observed with photodetectors. Thus, its determination can be accomplished only via an indirect measurement, i.e., using wavefront sensing, which maps the phase information in an imaginary variable to an intensity signal that can be measured with photodetectors.

Current wavefront sensing techniques fall into two categories, that is, microlens-array focal plane imaging and interferometric measurement approaches [4–11]. Both approaches have intrinsic limitations: the former decreases the system spatial resolution and phase sensitivity, and the latter requires multistep interference and phase shifting operations, making the system complicated and sensitive to vibrations. These limitations cause many applications to abandon the complicated but important characterization of SCA. The absence of wavefront information has obvious adverse impacts on the performance and procedure simplicity of light field characterization. Taking the best-known beam quality characterization as an example [12,13], to characterize the beam quality without wavefront information (because the commonly used beam profiler can only record the intensity profile), it is necessary to record a series of intensity profiles of the beam in different propagation planes. However, even so, due to the lack of full SCA information, it is still impossible to predict the exact diffraction behavior of a beam in free space or an optical system.

In this work, we present a complex-amplitude profiler capable of determining the exact SCA of a light field with a single-shot measurement. This novel technique exploits spatially resolved polarimetry (or Stokes tomography) with ancillary light, which is realized by using low-cost, integratable polarization sorting elements and a commercial camera. Below, we experimentally show that, with the SCA determined in a single propagation plane, one can achieve the full characterization of the measured beam, including the whole diffraction behavior and corresponding modal constitution.

## Principle and Methods

To prove the principle with specific examples, here, we chose paraxial structured Gaussian beams as unknown signals to be characterized. The transverse SCA of a beam upon propagation (referred to as the complete 3D spatial wavefunction) can be fully described (or predicted) by a solution of the paraxial wave equation, an analog of the Schrödinger equation for free particles with time $t$ replaced by the propagation coordinate $z$, expressed as

$$\psi_{\text{sig.}}(\boldsymbol{r}, z) = u(\boldsymbol{r}, z) \exp[iv(\boldsymbol{r}, z)], \tag{1}$$

where $u(\boldsymbol{r}, z)$ and $v(\boldsymbol{r}, z)$ are the spatial amplitude and wavefront, respectively, and $\boldsymbol{r}$ denotes the transverse coordinates. In particular, Eq. (1) belongs to an eigensolution of the paraxial wave equation (commonly referred to as a paraxial eigenmode) when the beam profile on $z$ is self-similar; otherwise, it can be represented as a superposition of eigensolutions with different modal orders [14]. Notably, the full 3D spatial wavefunction can be retrieved from a transverse SCA in an arbitrary propagation plane using the diffraction (or path) integral [3,15]. Namely, to fully characterize a light beam, it is required to record both $u(\boldsymbol{r})$ and $v(\boldsymbol{r})$ exactly at the same measuring plane $z_m$.

For this task, as shown in Fig. 1(a), we introduce an ancillary beam, denoted as $\psi_{\text{anc.}}(\boldsymbol{r}, z)$, and combine it with the signal beam, both carrying orthogonal circular state of polarization (SoP) $\hat{e}_\pm$, to generate a vector spatial mode given by [16–18]

$$\Psi(\boldsymbol{r}, z) = \sqrt{a}\psi_{\text{anc.}}(\boldsymbol{r}, z)\hat{e}_+ + e^{i\theta}\sqrt{1-a}\psi_{\text{sig.}}(\boldsymbol{r}, z)\hat{e}_-. \tag{2}$$

Here, $a \in [0,1]$ is a weighting coefficient that controls the degree of spin-orbit nonseparability (also known as concurrence) at $\psi_{\text{sig.}}$-$\psi_{\text{anc.}}$ overlapping positions, which is maximum for the position $\boldsymbol{r}_0$ as $\sqrt{a}g(\boldsymbol{r}_0) = \sqrt{1-a}u(\boldsymbol{r}_0)$; $\theta$ is the intramodal phase between the two polarization components. For convenience we take a well-known ancillary mode, the fundamental Gaussian mode, whose minimum beam width, acquired for $z = 0$, coincides with the measuring plane. That is, $\psi_{\text{anc.}}(\boldsymbol{r}, z_m) = g(\boldsymbol{r}) \exp[ikz]$, where $g(\boldsymbol{r})$ denotes a 2D Gaussian amplitude envelope whose cross section can well cover the signal mode. In this way, the vector mode at the measuring plane can be further represented as

$$\Psi(\boldsymbol{r}, z_m) = \sqrt{a}g(\boldsymbol{r})\hat{e}_+ + e^{i[v(\boldsymbol{r}, z_m)]}\sqrt{1-a}u(\boldsymbol{r}, z_m)\hat{e}_-. \tag{3}$$

Here, we assumed the global $\hat{e}_\pm$ intramodal phase $\theta = 0$ for simplicity, which can be realized in experiments via SoP control. In mode (3), the ancillary light works as a reference beam that copropagates with the signal along the same beam axis. Therefore, the vector mode can be regarded as a phase-stable polarizing interferometer.

We now perform a spatially resolved Stokes measurement on the vector mode given by Eq. (3), namely, $\langle\Psi(\boldsymbol{r}, z_m)|\hat{\sigma}_{1\text{-}3}(\boldsymbol{r})|\Psi(\boldsymbol{r}, z_m)\rangle$, where $\hat{\sigma}_{1\text{-}3}$ are Pauli matrices for SoP tomography [19-21]. The measured expectation values are a group of mutually unbiased observables named spatial Stokes distributions $S_{1\text{-}3}(\boldsymbol{r})$ (see Supplementary Materials for details), from which it is possible to reconstruct the spatially variant SoP of the vector mode. This measurement provides an in-situ visualization for vector modes generated and manipulated in experiments, so it has been widely used in the structured light community [22-26]. Remarkably, the complete information about the signal SCA is already hidden in the morphological feature of the observed spatially-variant SoP, which manifests in the spatial polarization orientation functions, defined as

$$\phi(\boldsymbol{r}) = \frac{1}{2}\text{atan2}\left(S_2(\boldsymbol{r}), S_1(\boldsymbol{r})\right), \tag{4}$$

where $\text{atan2}(y, x) \in [0, 2\pi)$ is the 2-argument arctangent. Since the function $2\phi(\boldsymbol{r})$ describes the relative phase between the two polarization components for each position in the beam cross section and since the ancillary wavefront is completely known, we can retrieve the wavefront of the signal from the relation $2\phi(\boldsymbol{r}) = v(\boldsymbol{r}, z_m)$. On this basis, combined with the signal amplitude $u(\boldsymbol{r}, z_m)$ that has been obtained in the SoP tomography (projection result on $\hat{e}_-$), the complete information about the signal beam at the measuring plane, i.e., $\psi_{\text{sig.}}(\boldsymbol{r}, z_m)$, is determined.

Regarding the accuracy (or uncertainty) of the wavefront determination, in this technique, the wavefront information of the beam is first converted to the intramodal phase of the vector mode in Eq. (3), reflected in the spatial polarization orientation in Eq. (4), and then determined via spatial Stokes tomography. That is, the accuracy of the wavefront estimation relies on that of the polarization-orientation estimation for each position (or pixel) in the beam cross section. From this perspective, the accuracy reaches its maximum as the local polarization reaches a linear SoP, i.e., as $\sqrt{a}g(\boldsymbol{r}_0) = \sqrt{1-a}u(\boldsymbol{r}_0)$ for the position $\boldsymbol{r}_0$, which corresponds to a minimum value (zero) in $S_3$, and gradually decreases as it approaches a circular SoP. Therefore, we can use the spatial concurrence of the observed spatially varying SoP, given by $C(\boldsymbol{r}) = \sqrt{1-S_3^2(\boldsymbol{r})} \in [0,1]$, to indicate the accuracy of the wavefront estimation [19, 27], as shown in the schematically in Fig. 1(a). For a region with a near circular SoP, i.e., $\left|\sqrt{a}g(\boldsymbol{r}_0) - \sqrt{1-a}u(\boldsymbol{r}_0)\right| \approx 1$, one can adjust the power of ancillary light (i.e., $a$) to make the local concurrence $C(\boldsymbol{r}_0)$ approach 1 and remeasure the SoP distribution there.

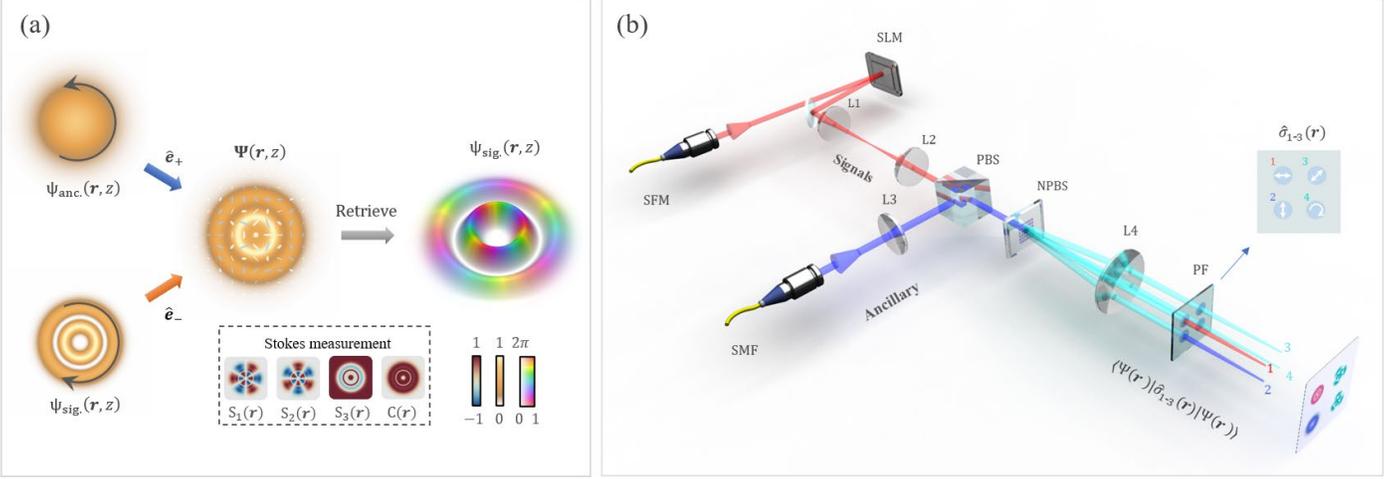

**Figure 1.** (a) Schematic of the principle. (b) Experimental setup, where the key components are the single-mode fiber (SFM), polarizing beam splitter (PBS), nonpolarizing beam splitter (NPBS), polarizing filter (PF), spatial light modulation (SLM), and lens (L1-L4).

## Experimental Results

Figure 1(b) shows the schematic setup for the proof-of-principle demonstration. A tunable continuous laser operating at a wavelength of 780 nm was used as the light source. We began with a $TEM_{00}$ mode collimating from a single-mode fiber for easy control in further SCA shaping. In signal preparation, we used a spatial light modulator (SLM), in which we loaded a specially designed hologram for performing complex-amplitude modulation and digital propagation [28], to shape the signal light (represented by a red beam in the schematic setup) into a desired spatial mode with a controllable diffraction distance. The prepared signal was first relayed by a $4f$ imaging system (L1 and L2) and combined afterwards with the ancillary beam via a polarizing beam splitter (PBS) to obtain the vector mode discussed above. The ancillary beam came from the same laser source and was shaped into also a $TEM_{00}$ mode by collimating from a single-mode fiber. To avoid the challenge that comes from SoP intramodal-phase jitter, the spatially-variant SoP of the vector mode was measured in a single-shot manner. Specifically, the vector mode was divided into 4 equal parts, which were then projected into horizontal, vertical, diagonal, and circular SoPs. After that, the four SoP projections, which can be converted to $S_{0-3}(\boldsymbol{r})$, were recorded simultaneously within the same frame of the camera (not drawn in the figure). Here, a nonpolarizing beam splitter (NPBS) with four output ports was realized by a liquid-crystal geometric phase, and the polarizing filter (PF) was comprised by waveplates and a polarizer (see Supplementary Materials for details). Moreover, lens L4 performed a Fourier transformation for the signal beam in combination with the SLM to realize digital propagation; the ancillary beam was expanded 2 times by lenses L3 and L4 to ensure that it could fully cover the signal beam.

In experiments, we first took the well-known structured Gaussian modes—the Hermite–Laguerre–Gaussian (HLG) family—as signal beams to be characterized [29]. Figure 2(a) shows the observed SCA patterns of the prepared signal beams at the waist plane ($z = 0$). The patterns in the top and bottom rows are Laguerre–Gaussian (denoted as $LG_{\ell,p}$) and Hermite–Gaussian (denoted as $HG_{m,n}$) modes, respectively, and their intermediates are shown in the middle row. The three HLG modes in each column correspond to a continuous astigmatic transformation from the LG polar to the HG equator on the same spatial-mode sphere with the same order $N$, where $N = 2p + |\ell| = m + n$ and $\ell = m - n$ [30–32]. Note that, except for the special case $N > 1$, the intermediate HLG mode can no longer be represented as a linear superposition of the two orthogonal LG (or HG) modes on the sphere but as a superposition of all the possible LG (or HG) modes with an order of $N$ (i.e., including modes not on this sphere) [14]. All the measurements agree well with the theoretical SCAs of the desired modes; see extended data in the Supplementary Materials. These SCAs provide complete knowledge about the spatial structure of measured beams, with which, in principle, we can exactly predict the diffraction behavior of each beam in free space or an optical system. That is, we can theoretically reconstruct the full spatial wavefunction of a beam, i.e., $\psi_{\text{sig.}}(r, z)$, by using the diffraction integral with the SCA observed at the measuring plane, i.e., $\psi_{\text{sig.}}(r, z_m)$, as the source (or pupil function); see Supplementary Materials for details.

Since the modes in Fig. 2(a) all belong to eigensolutions of the paraxial wave equation, their beam profiles should be propagation invariant apart from an overall change in size. To validate the observations in Fig. 2(a), we compare the changes in beam diameter upon diffraction obtained from the theoretical reconstruction and the slice-by-slice observation. This characterization method also relates to the generalized beam quality measurement for high-order Gaussian beams, i.e., the $M^2$ value and the ideal value is $M^2 = N + 1$ [33]. Figure 2(b) provides a visual description of the method, with the $\text{LG}_{2,1}$ mode as the example, in which the two 3D beam profiles used for the comparison are obtained via slice-by-slice observation (using digital propagation technique) and theoretical reconstruction (using diffraction integral). The similarity between both makes difficult to determine the difference between the two 3D beam profiles with naked eye. Figure 2(c) further provides an intuitive comparison for all five LG modes, where all the observed beam sizes (dots) fit closely on their corresponding predictions (curves) calculated through numerical diffraction, confirming the observation accuracy in Fig. 2(a). Thus, not surprisingly, the beam quality factors ($M^2$ values) calculated with the two approaches (dots and curve) as input are almost the same. Notably, beam characterization based on the $M^2$ measurement has many limitations for high-order modes, especially for superposition modes.

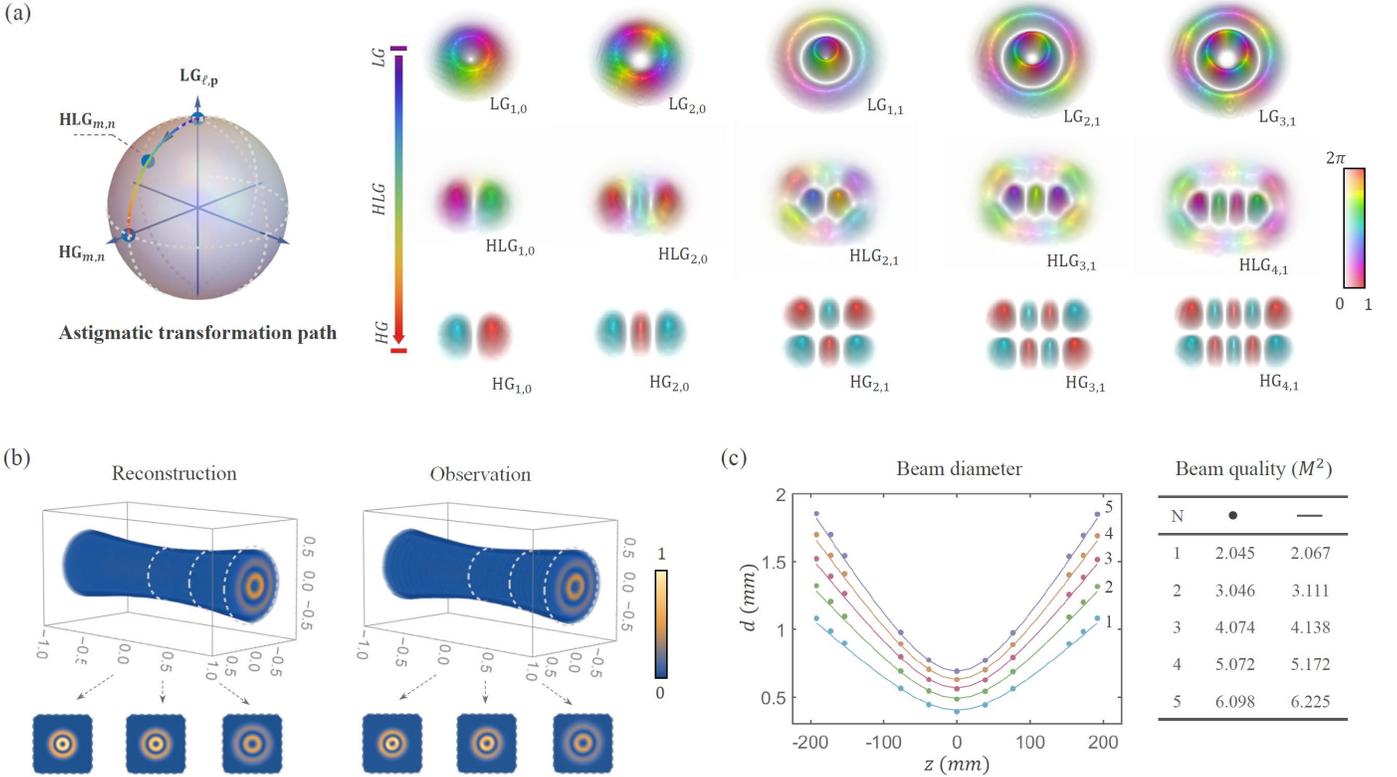

**Figure 2** Experimental results for paraxial eigenmodes. (a) Observed HLG modes with orders $N$ from 1 to 5, where the colored arrow displays the path of the astigmatic transformation from LG to HG modes on the modal sphere; (b) visual comparison for numerically reconstructed (diffraction integral) and experimentally observed (via digital propagation) 3D profiles of the measured $\text{LG}_{2,1}$ mode; and (c) comparison of observed and predicted changes in beam size of measured LG modes, as well as associated $M^2$ values.

We further proved the principle by characterizing the beam structure evolution of two superpositions of LG modes, given by $c_{0,0}\text{LG}_{0,0} + c_{4,0}\text{LG}_{4,0}$ and $c_{1,0}\text{LG}_{1,0} + c_{1,2}\text{LG}_{1,2}$, and here, we set all complex coefficients $c_{\ell,p} = \sqrt{1/2}$ for simplicity). Owing to their different Gouy phases $(N+1)\arctan(z/z_R)$, the beam structures would no longer be propagation invariant. Relevant studies have indicated that the changes in beam structure manifest in azimuthal rotation and radial oscillation and both exhibit pattern revival in the far field [34,35]. This inference was confirmed by the results in Figs. 3(a) and (b), which show the beam structure evolution of the two superposed LG modes. That is, $\psi_{\text{obs.}}(r, z)$ is recorded at the $z = 0$, $z = z_R$, and $z = z_\infty$ (Fourier) planes, and $\psi_{\text{num.}}(r, z)$ is calculated with $z = z_R$ and $z = z_\infty$ by using the observation at $z = 0$ as input. The numerical reconstruction looks the same as the measured again. To make a quantitative examination, we calculated the fidelity of reconstruction, defined by $F = \langle \psi_{\text{obs.}}(\boldsymbol{r}, z) | \psi_{\text{num.}}(\boldsymbol{r}, z) \rangle$, and the high fidelity shown in Fig. 3(c) validates the accuracy of the characterization.

Finally, beyond morphological observations, we can further derive the modal constitution of light fields (i.e., quantum state or density matrix) in terms of spatial degrees of freedom. This task is usually accomplished by using quantum state tomography [19,20], which requires multiple projections on all mutually unbiased bases, leading to a complicated and time-consuming operation, especially for high-dimensional states. Moreover, recent works showed that astigmatic operation and machine-learning based pattern recognition can be also used to determine simple OAM superimposition states [36,37]. In contrast, the SCA characterization provides complete information on light fields and thus enables to directly and easily obtain the complex-valued modal spectrum by computing projections on a complete orthogonal basis, such as the LG basis set; see Supplementary Materials for details. This capability has great potential application in high-dimensional state characterization and optical topology [19, 38–44]. Additionally, this enables to conveniently analyze and predict the beam evolution of the light fields mentioned above from an algebraic viewpoint [34,35]. For instance, Figures 3(d) and (e) show the computed modal constitutions (i.e., the LG modal complex spectra $c_{\ell,p}$) of the two superposed modes, respectively. It was shown that the modal weights (or power spectrum) remained constant upon propagation, i.e., $c_{\ell,p}^2 = 1/2$, while the intramodal phase between two successive LG components, due to unsynchronized Gouy phase accumulation, varied from 0 to $\pi$ when arriving $z = z_R$ and finally backed to 0 (or $2\pi$) at the far field $z = z_\infty$. This analysis clearly shows the mechanism of beam evolution and revival upon diffraction from an algebra perspective, which in fact exists widely in compounded systems undergoing unitary transformations [44–47]. In addition, modal constitutions of HLG modes in Fig. 2(a) are provided in the Supplementary Materials, showing that each mode has an identical modal order and thus is propagation invariant.

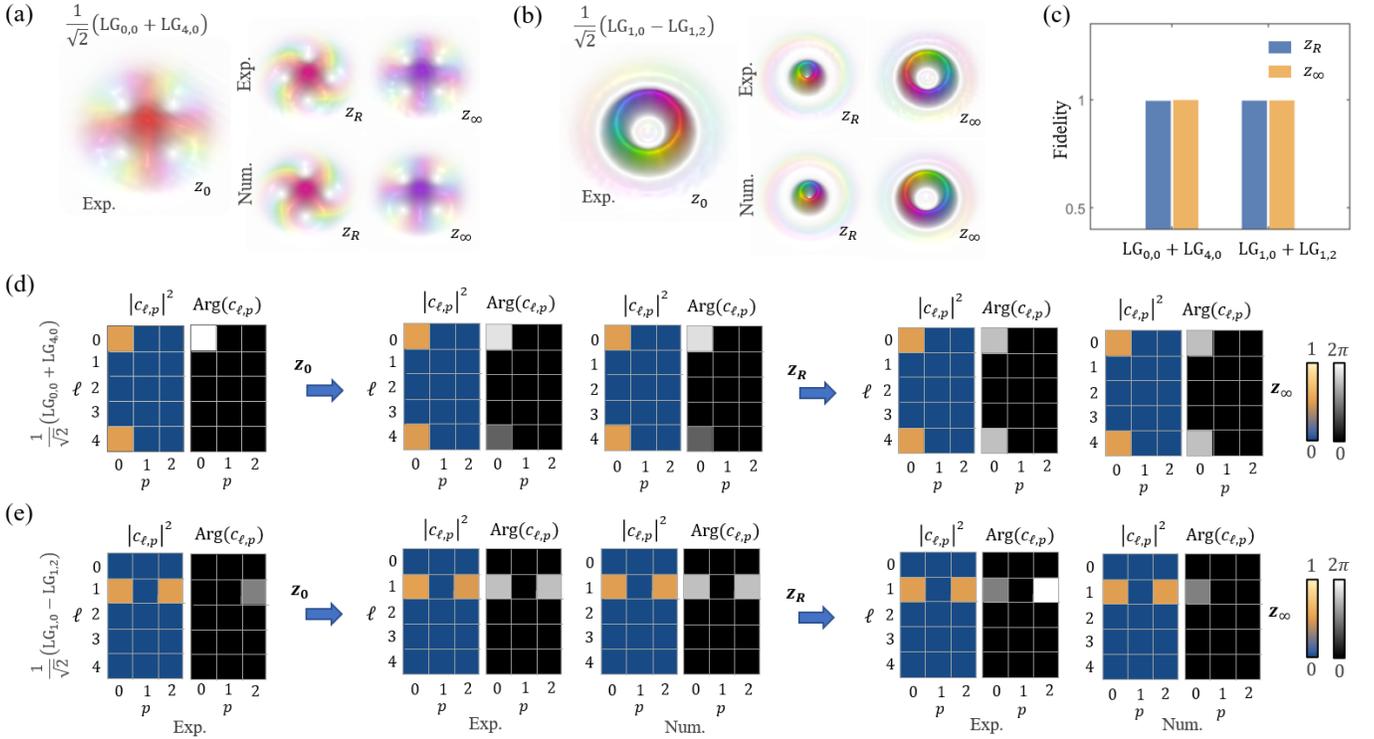

**Figure 3** Experimental results for noneigen paraxial modes. (a), (b) Observed superposed LG modes at $z = 0$, $z = z_R$, and $z = z_\infty$ (Fourier) planes, and corresponding numerical reconstructions based on observations at $z = 0$ plane; (c) SCA reconstruction fidelity relative to the observation; and (d), (e) modal constitutions and of two superposed LG modes and associated unitary evolution upon propagation.

## Conclusion

We have experimentally demonstrated a 'complex-amplitude profiler' that can completely determine the SCA of a light field with single-shot measurement. This novel technique exploits spatially resolved polarimetry with ancillary light to determine both the amplitude and wavefront of unknown beams. Specifically, the unknown signal light field to be measured was first combined with a preknown ancillary light via orthogonal polarization states, and forming a vector mode equivalent to a phase-stable polarizing interferometer; then, by performing a spatial Stokes measurement on the vector mode, the SCA of the signal can be determined with high precision and spatial resolution. On this basis, one can further theoretically reconstruct the full 3D spatial wavefunction and derive the modal constitution in the spatial degrees of freedom exactly. In particular, the device is capable of performing a single-shot measurement, providing it with real-time characterization and insensitive to vibrations. In addition, this technique only requires some low-cost and integratable polarization sorting elements that can be easily incorporated into the existing beam characterization system. Notice, however, that the ancillary light and the measured beam should be come from the same laser source (for coherent states) or photon (for single-photon states). A feasible way in experiment is using a single mode fiber to sample partial energy (or probability amplitude) of the signal light (or photons).

Since SCA characterization allows to access complete information of light fields, the principle demonstrated here has a broad spectrum of potential applications in various areas of optics. For instance, this principle can help researchers to unambiguously probe the topology hidden in a structured light field or the high-dimensional quantum state encoded with spatial degrees of freedom [38–44]. In addition to fundamental studies, this principle can also be coupled with developing optical techniques based on structured light, such as in situ detection on spatially multiplexed channels or structured sensing beams [39,48]. Beyond paraxial beams, this technique can also be used to completely characterize the complex spectrum of an imaging signal, enabling further implementation of arbitrary spatial analog computing on the signal [49].

## Acknowledgements

This work was supported by the National Natural Science Foundation of China (Grant Nos. 11934013, 62075050, 61975047, and 62275070), the High-Level Talents Project of Heilongjiang Province (Grant No. 2020GSP12), and the Heilongjiang Provincial Education Department (Grant No. UNPYSCT-2020185).

implementation with intense light: A tutorial. Advances in Optics Photonics 11: 67-134 (2019).
20. D'Ariano, M.-G. Paris and M.-F. Sacchi. Quantum Tomography. Advances in Imaging and Electron Physics 128: 205-308 (2003).
21. A. Selyem, C. Rosales-Guzmán, S. Croke, A. Forbes and S. Franke-Arnold. Basis-independent tomography and nonseparability witnesses of pure complex vectorial light fields by Stokes projections. Physical Review A 100: 063842 (2019).
22. A. Forbes, M. d. Oliveira and M.-R. Dennis. Structured light. Nature Photonics 15: 253 (2021).
23. J. Wang, F. Castellucci and S. Franke-Arnold. Vectorial light-matter interaction: exploring spatially structured complex light fields. AVS Quantum Science 2: 031702 (2020).
24. B. Zhao, X.-B. Hu, V. Rodríguez-Fajardo, Z.-H. Zhu and C. Rosales-Guzmán. Real-time Stokes polarimetry using a digital micromirror device. Optics Express 27: 31087 (2019).
25. X.-Y. Zhang, H.-J. Wu, B.-S. Yu, C. Rosales-Guzmán, Z.-H. Zhu, X.-P. Hu, B.-S. Shi and S.-N. Zhu. Real-Time Superresolution Interferometric Measurement Enabled by Structured Nonlinear Optics. Laser Photonics Reviews 2200967 (2023).
26. C.-Y. Li, S.-J. Liu, B.-S. Yu, H.-J. Wu, C. Rosales-Guzmán, Y. Shen, P. Chen, Z.-H. Zhu and Y.-Q. Lu. Toward Arbitrary Spin-Orbit Flat Optics Via Structured Geometric Phase Gratings. Laser Photonics Reviews 2200800 (2023).
27. B. Ndagano, H. Sroor, M. Mclaren, C. Rosales-Guzmán and A. Forbes. Beam quality measure for vector beams. Optics Letters 41: 3407-3410 (2016).
28. C. Rosales-Guzmán and A. Forbes. *How to shape light with spatial light modulators*.(SPIE Press, 2017).
29. E.-G. Abramochkin and V.-G. Volostnikov. Generalized Gaussian beams. Journal of Optics A: Pure and Applied Optics 6: S157 (2014).
30. R. Gutiérrez-Cuevas, M.-R. Dennis and M.-A. Alonso. Generalized Gaussian beams in terms of Jones vectors. Journal of Optics 21: 084001 (2019).
31. M.-W. Beijersbergen, L. Allen, H. E. L. O. v. d. Veen and J.-P. Woerdman. Astigmatic laser mode converters and transfer of orbital angular momentum. Optics Communications 96: 123-132 (1993).
32. H.-J. Wu, B.-S. Yu, J.-Q. Jiang, C.-Y. Li, C. Rosales-Guzmán, S.-L. Liu, Z.-H. Zhu, B.-S. Shi, Observation of anomalous orbital angular momentum transfer in parametric nonlinearity. Phys. Rev. Lett. 130, 153803 (2023)
33. C. Schulze, D. Flamm, M. Duparré and A. Forbes. Beam-quality measurements using a spatial light modulator. Optics letters 37: 4687-4689 (2012).
34. R.-Y. Zhong, Z.-H. Zhu, H.-J. Wu, C. Rosales-Guzman, S.-W. Song and B.-S. Shi. Gouy-phase-mediated propagation variations and revivals of transverse structure in vectorially structured light. Physical Review A 103: 053520 (2021).
35. B. P. da Silva, V.-A. Pinillos, D.-S. Tasca, L.-E. Oxman and A.-Z. Khoury. Pattern revivals from fractional Gouy phases in structured light. Physical review letters 124: 033902 (2020).
36. B. Pinheiro da Silva, D. S. Tasca, E. F. Galvão, and A. Z. Khoury. Astigmatic tomography of orbital-angular-momentum superpositions. Phys. Rev. A 99, 043820 (2019).
37. B. Pinheiro da Silva, B. A. D. Marques, R. B. Rodrigues, P. H. Souto Ribeiro, and A. Z. Khoury. Machine-learning recognition of light orbital-angular-momentum superpositions. Phys. Rev. A 103, 063704 (2021).
38. M. Erhard, R. Fickler, M. Krenn and A. Zeilinger. Twisted photons: new quantum perspectives in high dimensions. Light: Science & Applications 7: 17146 (2018).
39. B. Ndagano, I. Nape, M. A. Cox, C. Rosales-Guzman and A. Forbes. Creation and Detection of Vector Vortex Modes for Classical and Quantum Communication. Lightwave Technol 36: 292-301 (2018).
40. D. Sugic, R. Droop, E. Otte, D. Ehrmanntraut, F. Nori, J. Ruostekoski, C. Denz and M.-R. Dennis. Particle-like topologies in light. Nature communications 12: 6785 (2021).
41. V. Salakhutdinov, E. Eliel and W. Löffler. Full-field quantum correlations of spatially entangled photons. Physical review letters 108: 173604 (2012).
42. F. Brandt, M. Hiekkamäki, F. Bouchard, M. Huber and R. Fickler. High-dimensional quantum gates using full-field spatial modes of photons. Optica 7: 98-107 (2020).
43. L.-W. Mao, D.-S. Ding, C. Rosales-Guzmán and Z.-H. Zhu. Propagation-invariant high-dimensional orbital angular momentum states. Journal of Optics 24: 044004 (2022).
44. Y.-J. Shen, B.-S. Yu, H.-J. Wu, C.-Y. Li, Z.-H. Zhu and A. V. Zayats. Topological transformation and free-space transport of photonic hopfions. Advanced Photonics 5: 015001 (2023).
45. B.-S. Yu, C.-Y. Li, Y.-J. Yang, C. Rosales-Guzmán and Z.-H. Zhu. Directly determining orbital angular momentum of ultrashort Laguerre-Gauss pulses via autocorrelation measurement. Laser Photonics Reviews 16: 2200260 (2022).
46. H.-J. Wu, L.-W. Mao, Y.-J. Yang, C. Rosales-Guzmán, W. Gao, B.-S. Shi and Z.-H. Zhu. Radial modal transition of Laguerre-Gauss modes in second-harmonic generation. Physical Review A 101: 063805 (2019).
47. J. Courtial, K. Dholakia, L. Allen and M.-J. Padgett. Second-harmonic generation and the conservation of orbital angular momentum with high-order Laguerre-Gaussian modes. Physical Review A 56: 4193-4196 (1997).
48. X.-Y. Zhang, H.-J. Wu, B.-S. Yu, C. Rosales-Guzmán, Z.-H. Zhu, X.-P. Hu, B.-S. Shi and S.-N. Zhu. Real-time superresolution interferometric measurement enabled by structured nonlinear optics. Laser & Photonics Reviews 2023

# Supplementary Document

## Single-shot, full characterization of the spatial wavefunction of light fields via Stokes tomography

Bing-Shi Yu[1], Hai-Jun Wu[1], Chun-Yu Li[1], Jia-Qi Jiang[1], Bo Zhao[1], Carmelo Rosales-Guzmán[2,3], Zhi-Han Zhu[1*], and Bao-Sen Shi[1,4*]

[1] Wang Da-Heng Center, Heilongjiang Key Laboratory of Quantum Control, Harbin University of Science and Technology, Harbin 150080, China
[2] School of Physics, Harbin Institute of Technology, Harbin 150080, China
[3] Centro de Investigaciones en Óptica, A.C., Loma del Bosque 115, Colonia Lomas del Campestre, 37150 León, Gto., Mexico
[4] CAS Key Laboratory of Quantum Information, University of Science and Technology of China, Hefei, 230026, China
* e-mail: zhuzhihan@hrbust.edu.cn and drshi@ustc.edu.cn

### 1. Hermite-Laguerre-Gaussian modes and modal conversion.

The transverse structure of a paraxial light field and its corresponding evolution upon propagation can be fully described by a solution of the paraxial wave equation (PWE), which can be regarded as an analog of the Schrödinger equation for free particles. Except for an ideal plane wave, any eigensolution of the wave equation corresponds to a physically realizable "eigen" spatial mode, whose transverse structures are propagation invariant apart from an overall change of size. For instance, the well-known Laguerre-Gauss (LG) and the Hermite-Gauss (HG) modes are eigensolutions of the PWE in cylindrical and Cartesian coordinates, respectively, and their complex amplitude are [1,2]

$$\text{LG}_{\ell,p}(r,\varphi,z) = \sqrt{\frac{2p!}{\pi(p+|\ell|)!}} \frac{1}{w(z)} \left(\frac{\sqrt{2}r}{w(z)}\right)^{|\ell|} \exp\left(\frac{-r^2}{w_z^2}\right) L_p^{|\ell|}\left(\frac{2r^2}{w_z^2}\right) e^{i\left[kz+\frac{kr^2}{2R_z}+\ell\varphi-(N+1)\phi\right]}, \qquad (S1)$$

$$\text{HG}_{m,n}(x,y,z) = \sqrt{\frac{2^{1-n-m}}{\pi n!\,m!}} \frac{1}{w(z)} \exp\left(\frac{-r^2}{w_z^2}\right) H_n\left(\frac{\sqrt{2}x}{w_z}\right) H_m\left(\frac{\sqrt{2}y}{w_z}\right) e^{i\left[kz+\frac{kr^2}{2R_z}-(N+1)\phi\right]}, \qquad (S2)$$

where $z_R = kw_0^2/2$, $w_z = w_0\sqrt{1+(z/z_R)^2}$, and $R_z = z^2 + z_R^2/z$ are the Rayleigh length, beam waist, and radius of curvature, respectively. $L_p^{|\ell|}(\cdot)$ and $H_{m/n}(\cdot)$ denote the Laguerre and Hermite polynomials, respectively. The term $\phi = \arctan(z/z_R)$ is the propagation-dependent Gouy phase; and their modal orders are defined as $N = 2p + |\ell| = m + n$. Above Hermite- and Laguerre-Gaussian mode families have their own special features in transverse structures but have the same fundamental $\text{TEM}_{00}$ mode, and each can independently constitute a unique and unbounded 2D Hilbert space. Therefore, in principle, we can represent any Gaussian mode by using coherent superpositions (i.e., complex series) of LG, HG, or their intermediate Hermite-Laguerre-Gaussian (HLG) modes. i.e., the generalized Hermite-Laguerre-Gaussian mode (HLG) mode. All the possible $N$-order HLG modes following the relation $\ell = m - n$ can form the surface of a unit sphere, called modal sphere [3,4], and unitary transformation for states on the sphere can be realized by using astigmatic operation.

Astigmatic transformation (AT) refers to a unitary operation based on an astigmatic propagation that makes a beam to have different Gouy phases with respect to a given orthogonal axes. Specifically, the Gouy phase of $\text{HG}_{m,n}$ can be represented as [4,5]

$$(N+1)\phi = (m+1/2)\phi_x + (n+1/2)\phi_y,$$
$$\phi_{x,y} = \arctan\left(\frac{\Delta z_{x,y}}{z_{Rx,Ry}}\right) \qquad (S3)$$

For instance, the modal conversion $\text{HG}_{1,0}^{45°} \to \text{LG}_{1,0}$ requires a relative phase retardation $\delta = \phi_y - \phi_x = \pi/2$. So that we have,

$$\text{HG}_{1,0}^{45°} = \sqrt{1/2}\,(\text{HG}_{1,0} + \text{HG}_{0,1}) \to \sqrt{1/2}\,(\text{HG}_{1,0} + i\text{HG}_{0,1}) = \text{LG}_{1,0}, \qquad (S4)$$

In order to check the accuracy in spatial mode generation and observation, Figure S1 shows the theoretical reference of observed HLG modes in Fig. 2 (a).

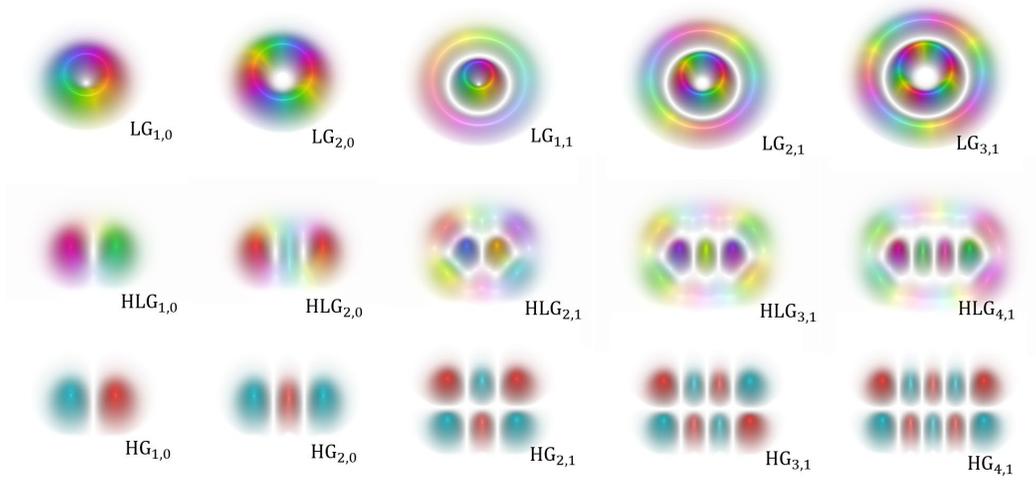

**Figure S1.** Extended data. Simulated spatial complex amplitudes at $z = 0$ of HLG modes shown in Fig. 2 (a).

In addition, the observed SCA profile at the measuring plane, denoted as $\psi(r, z_m)$, provides complete information about the light field. Thus, we can further retrieve the complete beam structure in 3D space, denoted as $\psi(r, z)$, by using Collins (or Kirchhoff) diffraction integral with $\psi(r, z_m)$ as the source [6,7], which in cylindrical coordinates can be expressed as

$$\psi(r,\varphi,z) = \frac{i}{\lambda z} exp(-ikz) \int r_0 dr_0 \int d\varphi_0 \, \psi(r_0,\varphi_0,z_m) \exp\left\{-\frac{ik}{2z} \times [r_0^2 - 2rr_0 \cos(\varphi-\varphi_0) + r^2]\right\}. \tag{S5}$$

## 2. Spatial-mode projection measurements enabled by spatial complex amplitude observation

Projection measurement in terms of spatial degrees of freedom has the same mathematical formalism with the Fourier transformation. For instance, the projection measurement to obtain complex-valued coefficients of a superposed LG mode $\psi(r) = \sum c_{\ell,p} \mathrm{LG}_{\ell,p}$ can be expressed as

$$c_{\ell,p} = |c_{\ell,p}| \exp[iArg(c_{\ell,p})] = \langle \mathrm{LG}_{\ell,p} | \psi(r) \rangle. \tag{S6}$$

Notably, the experimental implementation of such projective measurement, i.e., quantum state tomography, has been well established in optics. To date, however, there is no strict projective instrument for measuring spatial modes and instead, the widely used toolkit is the spatial autocorrelation measurement [8]. The far-field pattern of spatial autocorrelation function $C(r) = \psi(r)\mathrm{LG}^*_{\ell,p}(r)$ can in deed provide the real-value output $c^2_{\ell,p}$, which is thus equivalent to a standard experimentally projective measurement. However, owing to losing the imaginary part of $c_{\ell,p}$, one has to implement more projections in mutually unbiased basis of each two successive LG components. For the present technique, in contrast, because we have recorded the complete spatial complex amplitude (SCA) of the measured beam, $\psi_{\mathrm{obs.}}(r, z)$. Thus, we can directly access the complex-valued coefficients $c_{\ell,p}$ by performing Eq. (S6) with an observed $\psi(r)$ as input on a computer. Here, we provide also the calculated LG spectrum of observed HLG modes in Fig. 2(a), as shown below,

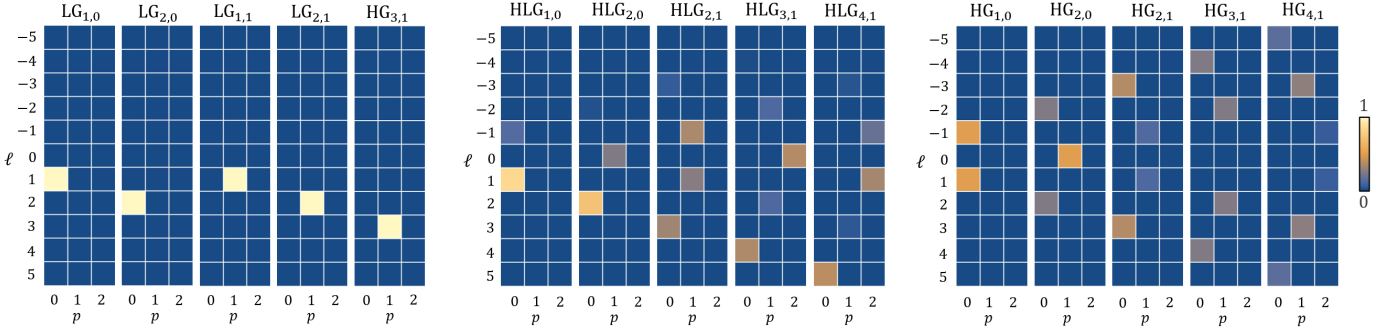

**Figure S2.** Extended data, calculated modal spectrum of observed HLG modes in Fig. 2(a)

### 3. Spatial Stokes tomography for vector modes.

The outputs of the spatial Stokes tomography mentioned in the main text, i.e., $\langle\Psi(r)|\hat{\sigma}_{0\text{-}3}(r)|\Psi(r)\rangle$, are a set of mutually unbiased real-valued functions [9]., or rather, spatial distributions of Stokes parameters, given by

$$\begin{cases} S_0(r) = |\sqrt{1-a}\psi_R|^2 + |\sqrt{a}\psi_L|^2 \\ S_1(r) = |\psi_H|^2 - |\psi_V|^2 \\ S_2(r) = |\psi_D|^2 - |\psi_A|^2 \\ S_3(r) = |\sqrt{1-a}\psi_R|^2 - |\sqrt{a}\psi_L|^2 \end{cases} \quad (S7)$$

where $\psi_{H,V,D,A,L,R}$ denote spatial probabilities or polarization-dependent spatial modes observed in the projections along the horizontal, vertical, diagonal, antidiagonal, left- and right-circular polarizations, respectively. Notice that $\psi_{H(D)}$ and $\psi_{V(A)}$ are not usually orthogonal to each other only in the case when $a = 0.5$. With the observations in Eq. (S7), we can in situ reconstruct the polarization resolved beam profile of the vector mode, expressed as

$$p(r) = S_0(r) \otimes \hat{\rho}(r). \quad (S8)$$

Here $S_0(r)$ describes the intensity profile of the beam, while $\hat{\rho}(r) = \frac{1}{2}\sum_{0\text{-}3} S_n(r)\hat{\sigma}_n$ is density matrix function that describes a spatially-varying polarization distribution covered on the beam profile, as shown in Fig. 1(a). According to Eqs. (7) and (8), the tomography can be further simplified into using only H, V, D and L polarizations projections. On this basis, in order to implement the simplified tomography within a single-shot measurement, we adopt a nonpolarizing beam splitter (NPBS) with four output ports and an integrated polarizing filter (PF), as shown below.

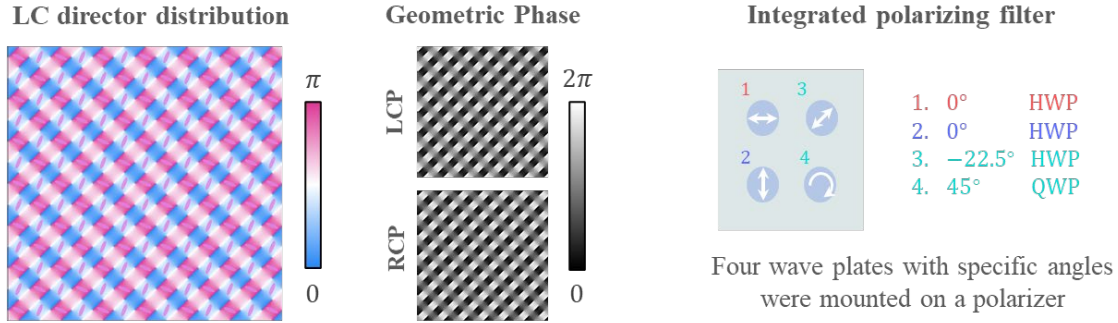

**Figure S3.** Details of the nonpolarizing beam splitter based on liquid-crystal geometric phase and the integrated polarizing filter.